# Unsteady loads on circular cylinder in cross-flow: an observation in natural wind and comparison with wind tunnel experiments

Yacine Manal[1] and Pascal Hémon[*]

LadHyX, CNRS-Ecole polytechnique, IP Paris, France

[1] PhD student at LadHyX

[*] corresponding author: pascal.hemon@ladhyx.polytechnique.fr

## Abstract

We present the synchronized unsteady wall pressure measurements performed at the circumference of a full scale 35 m circular chimney submitted to natural wind. The atmospheric turbulent boundary layer generates a turbulence intensity of 15 % at the altitude of the measurements and the corresponding Reynolds number is $1.13 \, 10^6$. For the first time, the steady and unsteady pressure distributions around the circumference are presented, despite large difficulties in the measuring process. The analysis of the pressure distribution using the bi-orthogonal decomposition (BOD) is a key factor in the success of the processing.

By comparison with wind tunnel similar measurements, with 2D cylinders, the time-averaged pressure distribution is found between the one of a subcritical regime and a supercritical regime, slightly closer to the subcritical regime notably because the base pressure coefficients are similar. The atmospheric turbulence is an essential ingredient for the unsteady pressure response on the chimney, which generates high levels of RMS pressure, much higher than all those measured in wind tunnel, even under the turbulent flow generated by a grid.

The analysis by BOD allowed the extraction of a term which is mainly generated by alternate vortex shedding. Despite a very noisy signal, the Strouhal number is determined to be 0.20-0.24. The corresponding shape of the pressure distribution is compared with those found in wind tunnel: it is shown that the full scale field measurements are close, but not identical, to the small scale measurements in wind tunnel on a rough cylinder submitted to a grid generated turbulent flow.

# 1. Introduction

## 1.1. Context of the research

As pointed out in (Demartino & Ricciardelli 2017) and references therein, the circular cylinder at various scales is very common in wind engineering problems, especially because of the alternate vortex shedding excitation. More specifically, large vertical slender flexible structures are subject to wind-induced vibrations involving complex inflow conditions due to the atmospheric boundary layer and high Reynolds number. Large structures with circular or mostly circular shapes, such as chimneys, stacks, wind turbine towers and launch vehicles, are responsive to alternate vortex shedding (Simiu & Scanlan 1978). However, there are a number of difficulties for studying such a phenomenon while respecting the natural wind characteristics such as the velocity gradient, the turbulence parameters and the Reynolds number similarity.

In such cases indeed, wind tunnel tests are performed on scaled models, which cannot respect the Reynolds similarity and introduce experimental difficulties. For instance, many authors add roughness elements on the surface of the model in order to simulate the high Reynolds number flow (Achenbach 1971; Szechenyi 1975; Barré & Barnaud 1995) although the technique is not perfectly validated (Ellingsen et al. 2022b).

The Strouhal number $St$, which represents the non-dimensional frequency of the alternate vortex shedding is strongly dependent on the Reynolds number and the upstream turbulence. It is defined as

$$St = \frac{f\,D}{\overline{U}} \qquad (1)$$

where the frequency is $f$, the diameter of the cylinder $D$ and the mean flow velocity $\overline{U}$. For a smooth circular cylinder it is well known (Blevins 2001; Chen 1987) that for subcritical Reynolds numbers, *ie* $Re \lesssim 200\,000$, the vortex shedding is well established and $St = 0.19 - 0.20$. In the critical region, $200\,000 \lesssim Re \lesssim 600\,000$ the vortex shedding becomes noisy or unstable, even temporarily absent, and the Strouhal number can be found in a wide region between 0.2 and 0.5. From these values of the Reynolds number, the Strouhal number is defined as the main frequency range of the flow excitation rather than a pure frequency. When the Reynolds number is further increased, $Re \gtrsim 600\,000$, the flow reaches the supercritical regime where one observe a re-organization of the wake, with an alternate vortex shedding having a Strouhal number subject to scattering, typically $St = 0.19 - 0.27$. This behavior is reported in the Figure 1 where the results from different authors are extracted from wind tunnel experiments.

Despite their interest, there are very few field studies of vortex excitation of chimneys that include wall pressure measurements and almost no one presenting unsteady data. In (Sageau 1978) and (Christensen et al. 1978) some wall pressure measurements were performed on some existing chimneys. An interesting experiment was presented in (Galemann & Ruscheweh 1992 ; Ruscheweh & Galemann 1996) where an experimental steel chimney of 28 m was equipped with a number of sensors, including wall pressure. In all these experiments, only the time-averaged wall pressures were measured, while unsteady data would be useful in the context of vortex shedding understanding in real wind conditions.

In 1983 few full scale data have been published from the observation of a stack in Australia (Melbourne et al. 1983). These pioneering results were obtained in stormy conditions and with the stack experiencing wind-induced excitation. Hence the knowledge of the Strouhal number in natural wind conditions with a motionless structure is still missing for practical applications.

For instance in wind engineering, it is common to have chimney with a typical diameter of 2 m and a natural first bending frequency of the order of 1 Hz. Then the critical wind velocity at which the resonance occurs with the alternate vortex shedding is around 10 m/s, which is a moderate value occurring very often in practice.

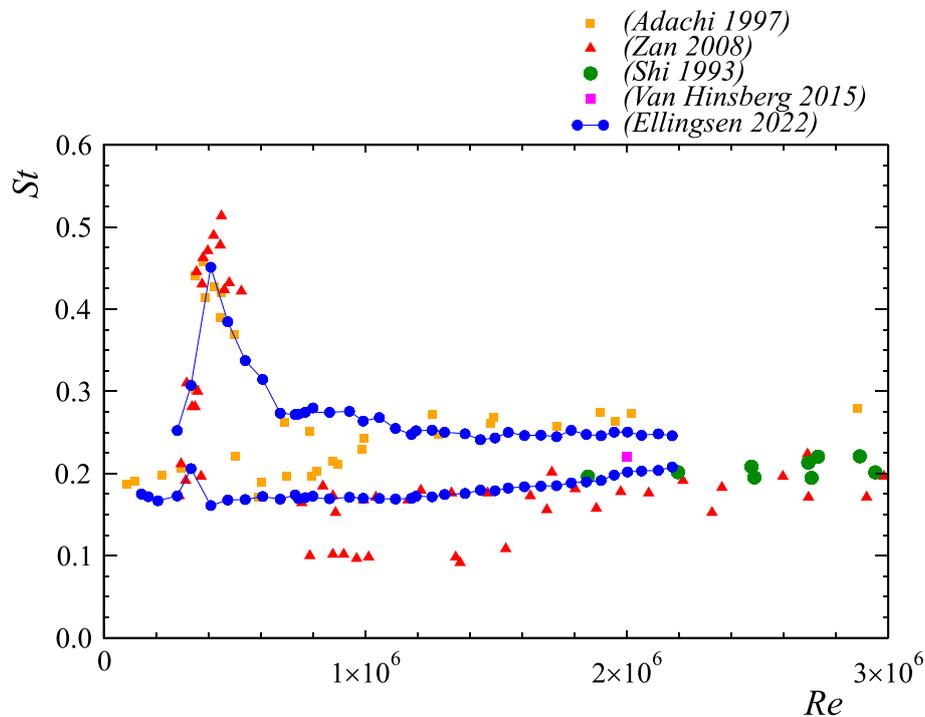

Figure 1. Strouhal number versus Reynolds number from different authors.

## 1.2. Content of the paper

In 2018 a project was started under a partnership including the company Beirens, the CNES, the CSTB and LadHyX. One of its components was the erection of an experimental chimney on an observation site for which first results were presented in (Ellingsen et al. 2021). In the current paper, we focus on the synchronized unsteady wall pressure distribution measured during a short term campaign in July 2021. Preliminary draft results have been communicated at the conference FIV2022 (Manal & Hémon 2022) but new findings and processed data are now presented.

The main part of the paper describes the experimental setup and the natural wind conditions, followed by the presentation of the pressure measurements on the circumference of the chimney. Next, these results are compared to the wind tunnel experiments in which similar measurements were previously conducted. One was performed in the large wind tunnel "Jules Verne" of CSTB in Nantes (Ellingsen *et al.* 2022a), with a large 2D cylinder in smooth flow and complete Reynolds similarity. The second experiment was at low scale with a 2D rough cylinder in a grid generated turbulent flow and was detailed in (Ellingsen *et al.* 2022b). The latter was conducted in the NSA wind tunnel of CSTB.

Views of these three experiments are shown in the Figure 2. The large cylinder in wind tunnel has as aspect ratio similar to the upper section of the chimney. However one must keep in mind the differences between these experiments, particularly the 3D versus 2D characteristics and the free top of the chimney versus the wind tunnel walls.

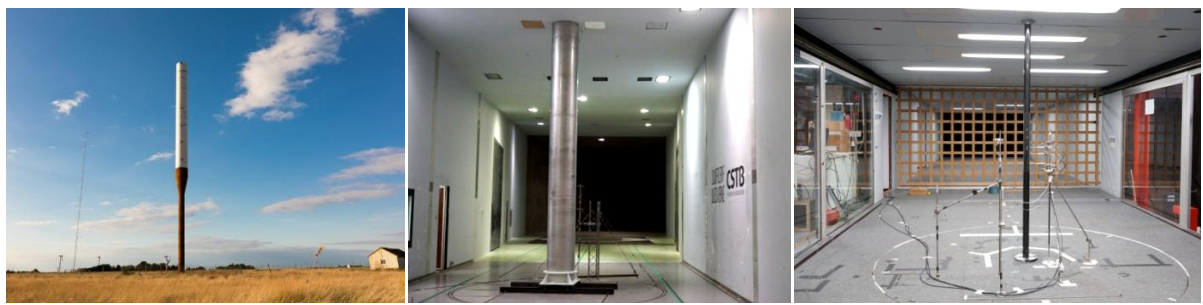

Figure 2. Views of the three experiments. From left to right: full scale chimney; large smooth cylinder in smooth flow; small rough cylinder in turbulent flow

## 2. The full scale experiment

The observation site is located in Bouin (85, Département of Vendée) in the western part of France, at about 2 km from the Atlantic seashore. The environment is a marsh which makes the relief very flat over a distance greater than 2 km around, as it can be seen in the Figure 3.

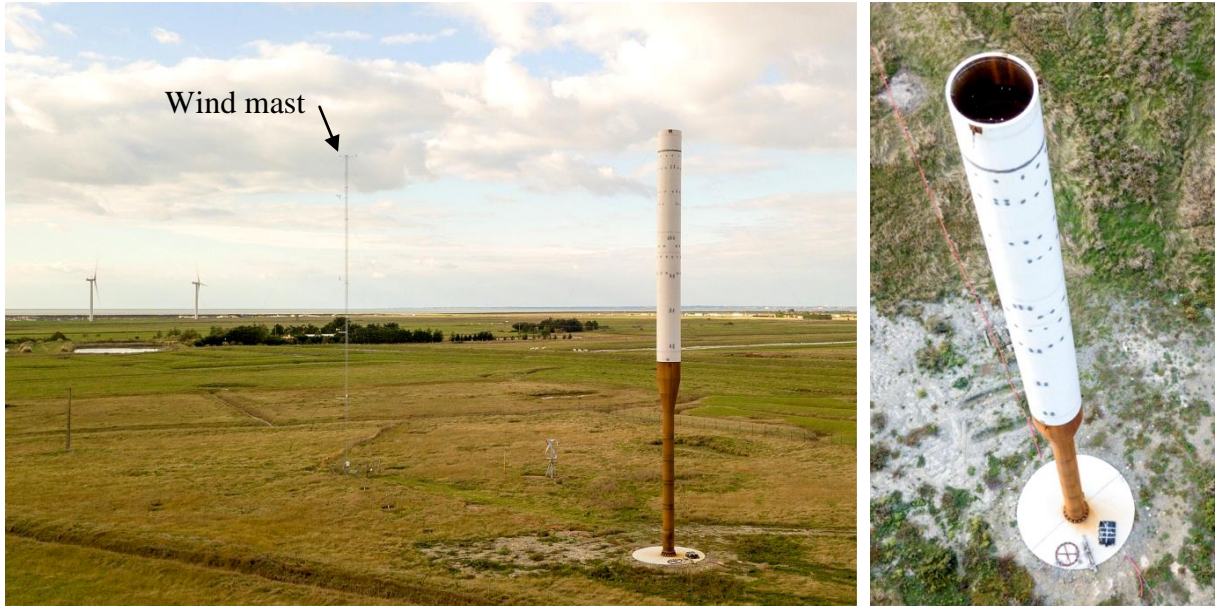

Figure 3. Views of the chimney and the wind mast.

### 2.1. Description of the experimental setup

The chimney is a steel tube with 35.5 m high and an external diameter of 1 m in the lower part, from 0 to 12 m, and 2 m in the upper part above 15 m. From 12 to 15 m the diameter linearly increases from 1 to 2 m. The upper part has an aspect ratio $L/D$ equal to 10. The chimney is clamped at the bottom in a concrete mass properly designed.

The purpose of this particular shape is to obtain a chimney with a low Scruton number and a first bending mode at a low frequency in order to get a lock-in with alternate vortex shedding at moderate winds. More details can be found in (Ellingsen et al. 2021) and we recall here the main useful information.

The total mass of the chimney is 11276 kg including all additional masses necessary for mounting, maintenance and human access. The equivalent mass $m_e$ per unit of height is computed using the first mode shape $\varphi(z)$ which is obtained via a structural analysis (Simiu & Scanlan 1978; Eurocode 2005):

$$m_e = \frac{\int_0^h \varphi(z)^2 m(z) dz}{\int_0^h \varphi(z)^2 dz} \qquad (2)$$

A value of $m_e = 322.6$ kg/m was obtained.

The Scruton number reads:

$$Sc = \frac{4\pi\eta\, m_e}{\rho D^2} \tag{3}$$

The reduced structural damping referred to the critical damping $\eta$ was measured in situ in two normal directions via the records of the chimney motion after a manual release. It was found that $\eta = 0.185 \pm 0.005$ %. Using the air density $\rho = 1.225$ kg/m³ and the upper diameter $D = 2$ m of the chimney, the Scruton number is determined to be $Sc = 1.53$.

The first bending frequency was measured in the same way as the reduced damping. It was found that $f = 0.868 \pm 0.001$ Hz.

The Strouhal number is supposed to be 0.18 in (Eurocode 2005) but potentially 0.21 (Ellingsen et al. 2022a). Therefore, the lock-in mean velocity is expected to be in the range $\bar{U}_c = 8.3 - 9.6$ m/s, a relatively moderate wind frequently observed on the site. Associated to the low Scruton number mentioned above, high amplitude oscillations at lock-in are expected. Therefore, the chimney is kept motionless most of the time by means of 3 stay cables. It is let free to oscillate in few sequences only under human monitoring.

## 2.2. Pressure measurements

The main data used in this paper are the synchronized wall pressure measurements. They are performed by using 32-channel pressure scanners (32HD ESP pressure scanners from Pressure Systems Inc.) with multiplex frequency of 70 kHz. The global accuracy is about ±1 Pa, but practical difficulties in setting the zero value (the no wind response of the sensor) lead to much higher errors on the mean component. So the classical time-averaged values of the pressure are not used and the mean pressure distribution will result from a data processing detailed hereafter. The records have a sampling frequency of 16 Hz and stored in sequences of 10 minutes long.

A number of taps have been mounted on the chimney but some of them were damaged by the environmental conditions. Fortunately the most interesting ones could be collected: these are the 32 taps around the chimney at 26.75 meters of height. They are located at mid-span of the 2 m diameter section, 10 m or $5D$ from the top and the diameter reduction. They are uniformly distributed around the circumference and spaced by 11.25° of the azimuth angle $\theta$.

The pressure coefficient is defined as

$$Cp(\theta, t) = \frac{P(\theta,t) - P_{ref}}{\frac{1}{2}\rho \bar{U}^2} \tag{5}$$

where $P(\theta, t)$ is the instantaneous measured pressure at the azimuth angle $\theta$. The reference pressure $P_{ref}$ is the mean static pressure inside the chimney at the same altitude of the external taps. It is obtained via the reference port of the pressure scanner which delivers directly the differential pressure $P(\theta, t) - P_{ref}$. $\rho$ is the air density corrected by the atmospheric pressure and the air temperature and $\bar{U}$ is the mean wind velocity.

The azimuth angle $\theta = 0°$ is referred to the wind direction as in a wind tunnel test section, so the data shown further in natural wind have been rotated in order to reach a "symmetrical" result which is of course imperfect due to the natural scatter of these observations. Moreover, despite the wind measurements (see below), the pressure coefficient was corrected "manually" by setting the parameter $\overline{U}$ so as to give a pressure coefficient $\overline{Cp} = 1$ at $\theta = 0°$. Practically, the process takes place in two steps: the first is to set the wind direction, and it was determined that the RMS distribution of the pressure coefficient is the most reliable data to reach a symmetrical distribution, thanks to the two sharp peaks on the sides of the cylinder. Having done this, it is then possible to adjust the mean wind velocity in equation (5) in order to get $\overline{Cp} = 1$ in the region facing the wind. Finally, the velocity was adjusted by 5 % by reference to the value measured at the wind mast, with 20° of rotation. Due to this empirical process, the uncertainty of the pressure distribution is larger than the raw uncertainty of the sensors and a rough estimation is 10 %.

## 2.3. The wind mast

A mast of 40 meters high is erected at the distance 55 meters from the chimney in the west direction. It is equipped with 4 anemometers at 10 (cup), 18 (propeller), 25 (3D sonic) and 35 (propeller) meters of height. Three wind vanes complement the cup and the propeller anemometers. All these sensors are shifted from the mast axis by 1.5 m in order to limit the interactions.

The cup anemometer at 10 m has an accuracy of ±0.1 m/s, while the propellers at 18 and 35 m have and accuracy of ±0.3 m/s. The vanes provide the wind direction with an accuracy of ±3°.

The 3D sonic anemometer has better characteristics, with an accuracy of ±0.05 m/s and ±2°. It continuously records the wind velocity components at the sampling frequency of 5 Hz. The 3 others anemometers record only the statistical values (mean, RMS, maxima) of the velocity modulus and its direction in degree, referred to magnetic North, over sequences of 10 minutes.

## 2.4. Wind measurements

Due to the instrumentation complexity and the necessity of having a proper weather, notably without rain, the observation and the measurements of the wall pressures occurred during two days, July 19-20 2021. Some interesting events have occurred during these two days with a mean speed around 8-10 m/s and a wind coming from North-East, typically 50-70° referred to magnetic North.

Finally, four sequences have been selected for the detailed processing and investigation. The two first from July the 19[th] are without motion of the chimney, while the two others from July the 20[th] are recorded during chimney oscillations. In the following, we will focus the results on the motionless sequence recorded the 19[th] of July at 4 PM. The second sequence at 4:10 PM, still without motion, gave the same result than the first one. The two other sequences the day after were recorded when the chimney was released of its stay cables and encountered oscillations that should be analyzed in a different way in another study. Moreover, there are reasonable doubts about the pressure measurements that could be influenced by the motion and therefore they are not presented here.

The atmospheric boundary layer is shown in the Figure 4 where the mean velocity and the turbulence intensity are presented versus altitude. They are compared with the Eurocode profiles for the roughness type II which is supposed to apply to the present site.

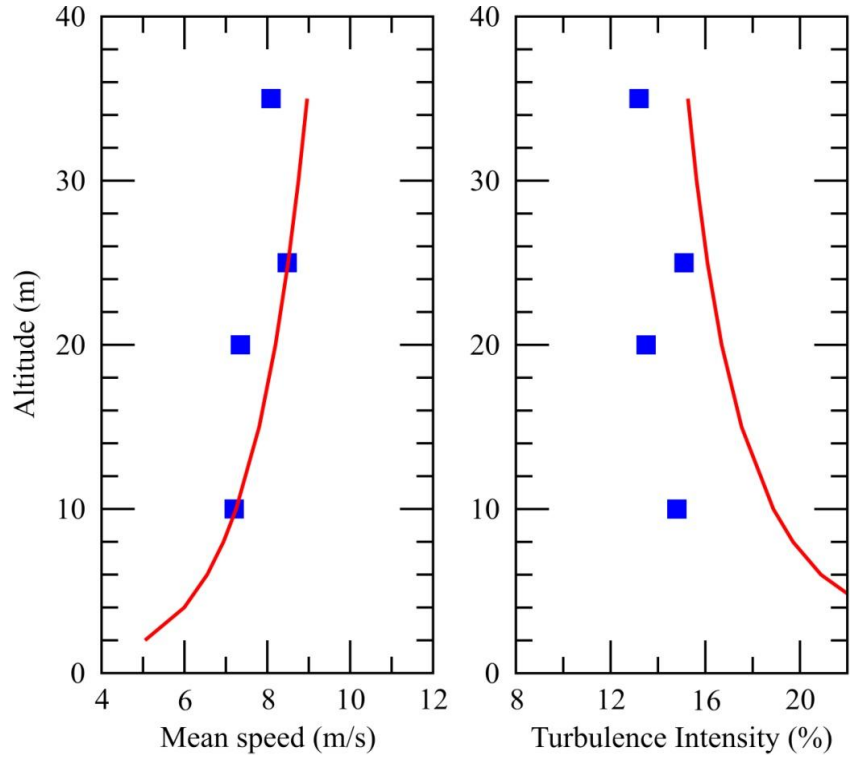

Figure 4. Mean velocities and turbulence intensities measured by the wind mast. Red line is the normalized profile prescribed by Eurocode. July 19th, 4:00 PM.

In such case, the mean velocity is given by

$$\bar{U}(z) = 0.19\, \bar{U}(10)\, ln\left(\frac{z}{z_0}\right) \tag{6}$$

and the turbulence intensity by

$$I(z) = 1/\, ln\left(\frac{z}{z_0}\right) \tag{7}$$

where the roughness height $z_0$ is 0.05 m (Eurocode 2005). One should note here that the mean speed at 10 m high is the reference speed for the type II roughness, so that every velocity profile should go through that point.

By looking at Figure 4, it appears that, while the mean velocity gradient follows more or less well the Eurocode profile, the measured turbulence intensity is lower than the one furnished by Eurocode. One explanation could be that the profile furnished by the standards is supposed to be in strong wind condition, typically with a return period of 50 years, while in the present case the wind is moderate.

Note that the sonic anemometer located at the altitude of 25 m is supposed to be more accurate for the turbulence measurement, by comparison with the other cup or propeller anemometers that have an inertial effect. The time evolution of the measurement of the sonic anemometer is shown Figure 5.

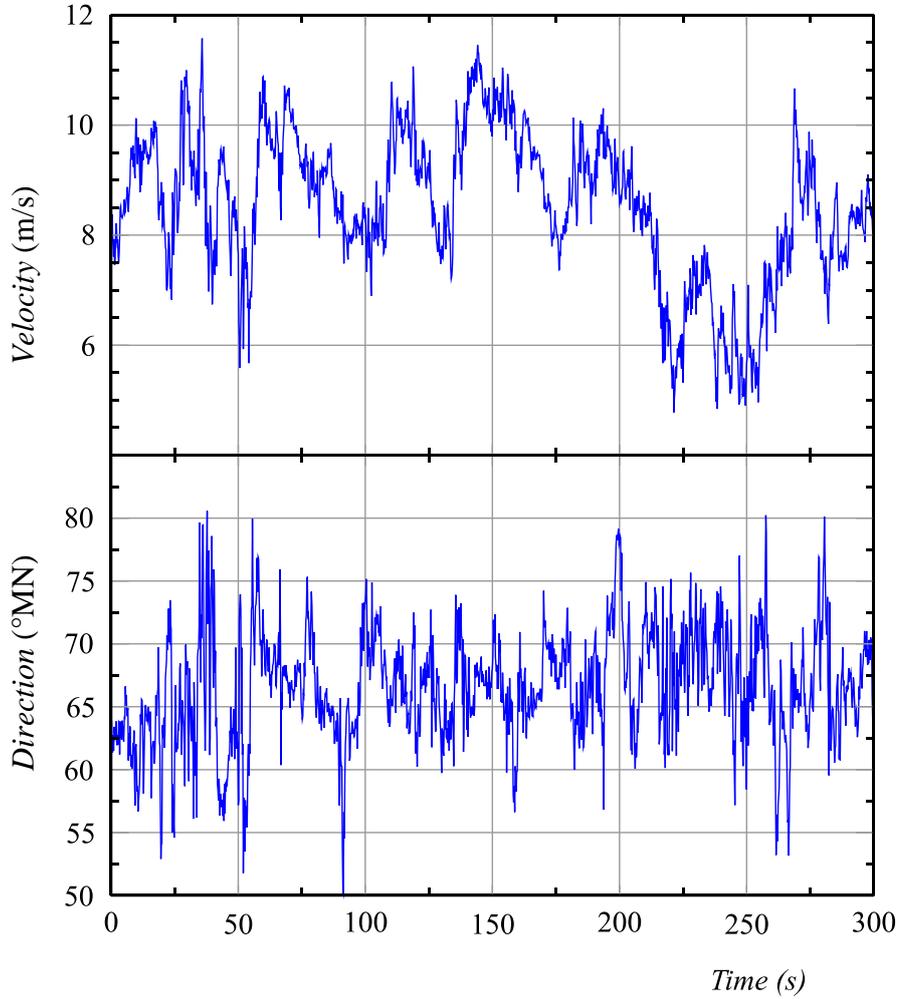

Figure 5. Time evolution of the wind velocity and direction of the selected sequence at 25 m, $\overline{U} = 8.49\ m/s$, $\sigma_u = 1.28\ m/s$, Direction 67°MN, $\sigma_{Dir} = 4.5°$, July 19th, 4:00 PM.

The power spectral density (PSD) of the longitudinal velocity is computed thanks to the time records furnished by the sonic anemometer. It can be compared to the Von Karman spectrum $S_u(f)$ which reads (Simiu & Scanlan, 1978):

$$\frac{S_u(f)}{\sigma_u^2} = \frac{4\ L_u}{\overline{U}\left(1+70.7\left(\frac{f\ L_u}{\overline{U}}\right)^2\right)^{5/6}} \qquad (8)$$

where $\sigma_u$ is the standard deviation of the velocity, $f$ the frequency and $L_u$ the integral scale of turbulence of the longitudinal component. The latter is determined by finding the best fit of the Von Karman model with the measured PSD, as shown in Figure 6 where $S_u(f)$ is normalized with $\sigma_u^2$ as in Eq. (8).

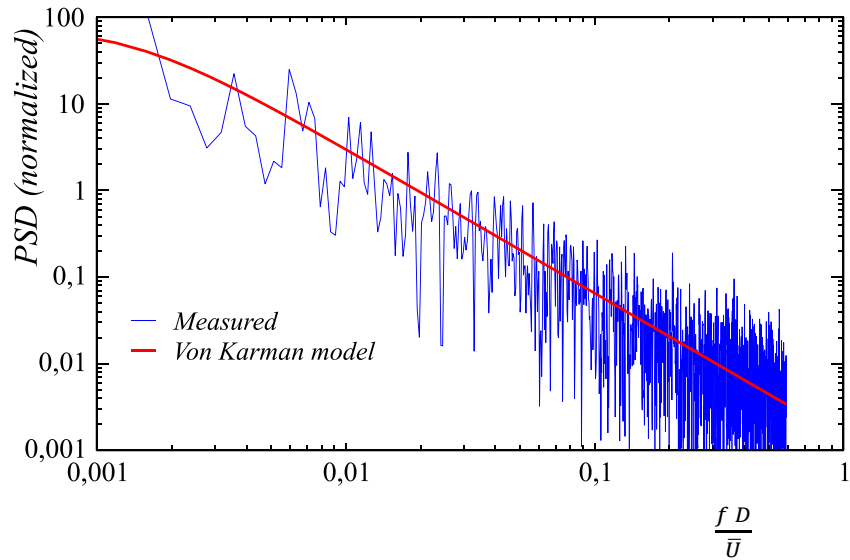

Figure 6. PSD of longitudinal velocity at 25 m, $\bar{U} = 8.49 \, m/s$, $\sigma_u = 1.28 \, m/s$, $L_u = 168 \, m$ ; Direction 67°MN, $\sigma_{Dir} = 4.5°$, July 19th, 4:00 PM.

## 2.5. Response of the chimney

Altough the wall pressure measurements presented below are the main topic of the paper, we give in the Table 1 the statistics of the two sequences obtained during the oscillations of the chimney. The amplitude $Y_{peak}$ is the peak to peak average amplitude of oscillations at the top, which were almost periodic at $f = 0.849$ Hz. The wind parameters are from the sonic anemometer at 25 m. These results show that the oscillations amplitudes are relatively small: the corresponding Strouhal number is 0.17 ±5%, suggesting that the "perfect" lock-in does not occur here.

Table 1. Ten minutes statistics of two sequences July 20[th] of the oscillating chimney

| Sequence | $\bar{U}$ (m/s) | $\sigma_u$ (m/s) | $L_u$ (m) | Dir (°) | $\sigma_{Dir}$ (°) | $Y_{peak}$ (m) |
|---|---|---|---|---|---|---|
| 12:40 | 9.96 | 1.31 | 50 | 67.4 | 8.1 | 0.14 |
| 12:50 | 9.72 | 1.03 | 50 | 73 | 7.6 | 0.15 |

## 3. Analysis of the wall pressure distribution

### 3.1. Sample results

Samples of pressure time histories are shown in the Figure 7. The signal from the tap facing the wind at $\theta = 0°$ presents slow and large variations of its amplitude which looks like those observed on the wind velocity previously shown in the Figure 5, remembering that the two measurements are spaced by 55 m.

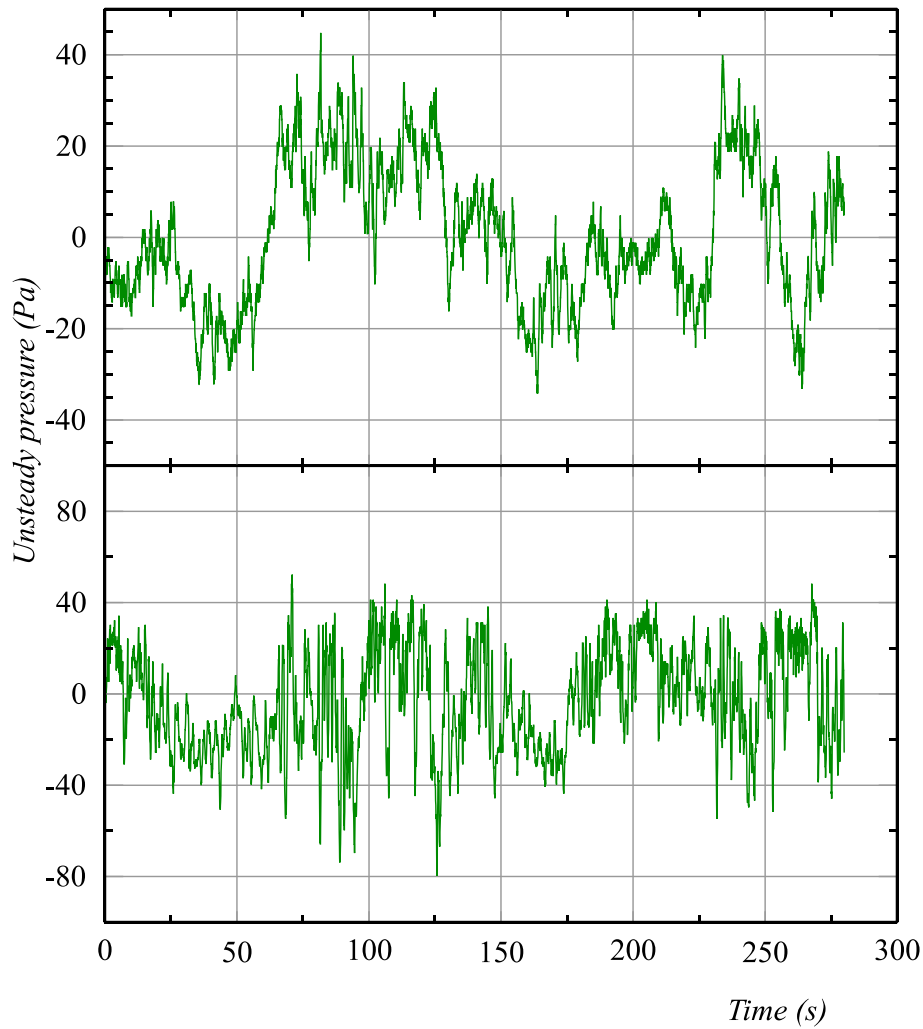

Figure 7. Centred pressure signals for taps at $\theta = 0°$ (top) and $\theta = 93°$ (bottom).

In contrast, the pressure measured on the side region of the chimney circumference at $\theta = 93°$ presents higher amplitudes of fluctuation, which seems more disconnected from the wind. But the Fourier analysis of that signal confirms its noisy content, where no remarkable frequency could be detected. Therefore, a more powerful analysis of the pressure measurements becomes necessary.

As said before, we present here the results obtained with one sequence because the three other sequences which were recorded gave similar results, notably two of the sequences with the chimney oscillating at low amplitude (Manal & Hémon 2022).

### 3.2. The bi-orthogonal decomposition (BOD)

We rapidly recall here the analyzing technique which was first introduced by (Aubry, Guyonnet & Lima 1991). The idea of the BOD is to decompose the spatio-temporal signal $Cp(\theta, t)$ in a series of spatial functions $\phi_i(\theta)$ named further as "topos", coupled with a series of temporal functions $\psi_i(t)$ named "chronos". The BOD can be written as

$$Cp(\theta, t) = \sum_{i=1}^{N} \alpha_i \, \phi_i(\theta) \, \psi_i(t) \tag{9}$$

where $\alpha_i$ are the eigenvalues of the spatial or the temporal covariance matrix of the signal $Cp(\theta, t)$. $N$ is the number of terms retained for the decomposition. Chronos and topos are orthogonal between them and normalized. Mathematical details can be found in (Aubry, Guyonnet & Lima 1991) and practical applications are presented in (Hémon & Santi 2003).

It was shown that the eigenvalues $\alpha_i$ are common to chronos and topos and that the series converge rapidly so that $N$ is possibly small compared to the original size $T$ of the problem. $T$ is the smallest value between the number of pressure taps and the number of time records, *ie* $T = 32$ in the present case. This means that the $\alpha_i$ have a numerical value that decreases rapidly. Their sum

$$A = \sum_{i=1}^{T} \alpha_i \tag{10}$$

represents the total energy in the original signal. Then each couple of chronos and topos have their contribution to the signal, which decreases as long as their rank $i$ increases. It is common to normalize the eigenvalues $\alpha_i$ with the total energy A, which will done in the analysis further.

Note that BOD is very similar to the so-called proper orthogonal decomposition (POD), except that the mean value of the original signal is kept in the analysis, refer to (Hémon & Santi 2003) for a discussion on that point.

### 3.3. Analysis of the chimney measurements

As mentioned before, the analysis by BOD is performed on the sequence measured the 19[th] of July at 4 PM. The sequence, initially of 10 minutes long, is cropped manually to one block of 280 s. We keep then a little less than 3 minutes, in order to get signals with "stationary" conditions. Finally, the analysis results from 32 pressure taps at the circumference of the chimney with 4480 time steps. Note that despite the large uncertainty on the mean value, due to the poor zero value calibration, no extraction of the time averaged value is performed prior to the analysis.

The BOD terms are presented hereafter in Figures 8 to 10. The eigenvalues are shown in Figure 8, the six first topos in Figure 9 and the corresponding chronos in Figure 10. Before all, one should notice that the term #5 is due to an erroneous pressure tap and has no physical meaning.

Despite their apparent complex shape, the 3 first topos are the components that makes the time-averaged pressure coefficient. These 3 terms represent 77 % of the total energy of the original signal. Following the BOD definition in equation (9), the combination of the time-averaged 3 first terms reads indeed

$$\overline{Cp(\theta)} = \sum_{i=1}^{3} \alpha_i \, \phi_i(\theta) \, \overline{\psi_i} \tag{11}$$

where the time-averaged chronos is denoted $\overline{\psi_i}$. The resulting time-average pressure coefficient $\overline{Cp(\theta)}$ is plotted in Figure 11. The curve is surprisingly smooth despite the mentioned errors during the measurements process. These random errors have been extracted by the BOD process and pushed in high rank terms associated to small eigenvalues. At $\theta = 0°$ the pressure coefficient is 0.9 instead of 1. By using a proper integration of that pressure coefficient, the mean drag is found to be $Cd = 0.79$.

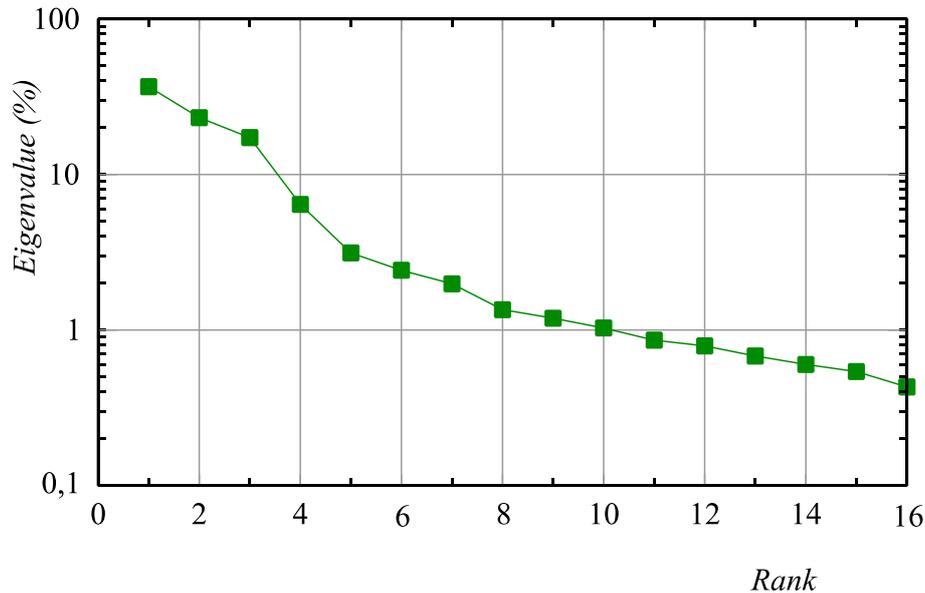

Figure 8. Eigenvalues of the BOD of the pressure distribution on the chimney.

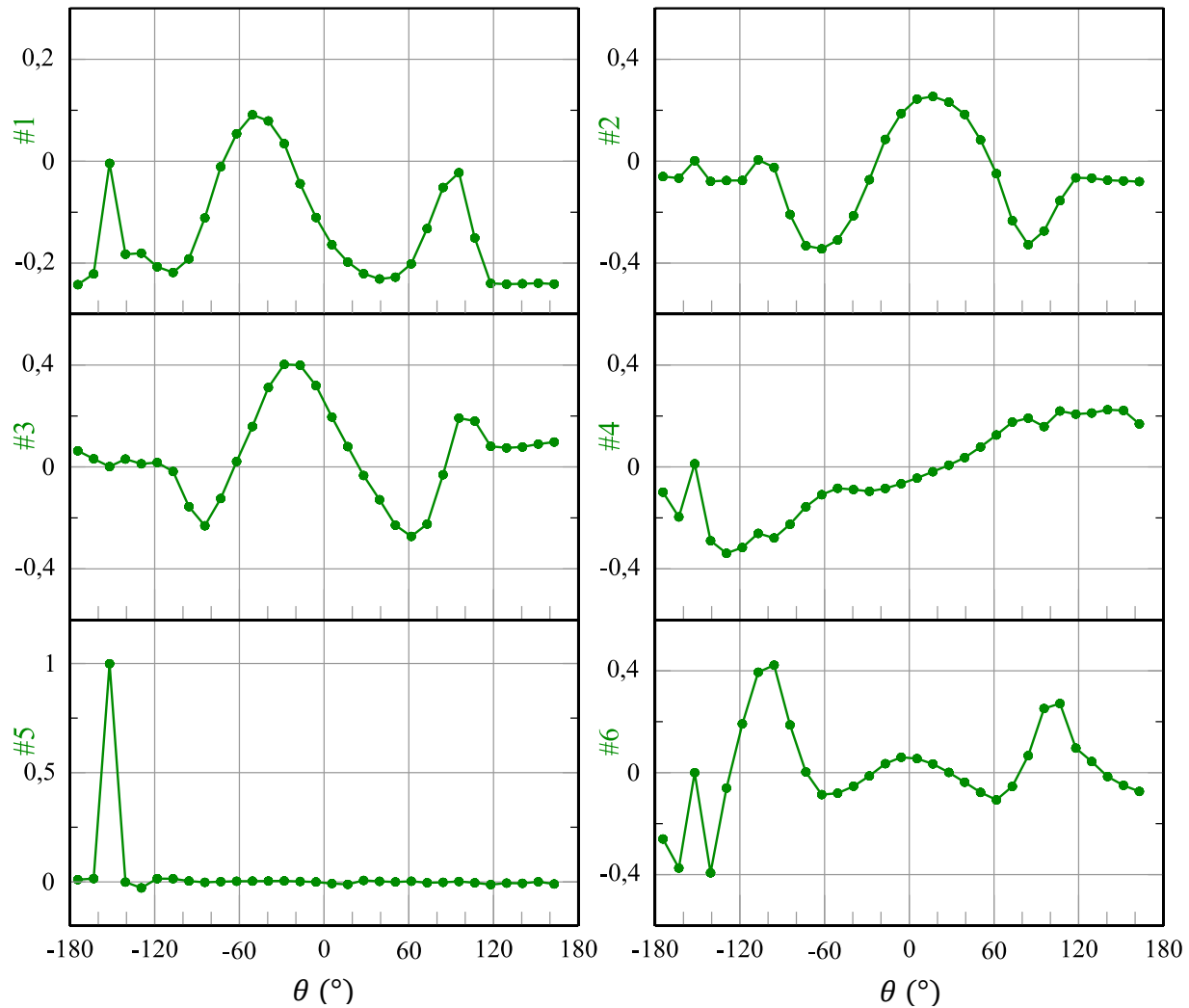

Figure 9. Six first topos of the pressure distribution on the chimney.

The term #4 represents the unsteady pressure distribution that is producing an unsteady lift force because of the anti-symmetric shape of the topos. It takes 6.4 % of the total energy but 28% of the unsteady part. Moreover, by making a Fourier analysis of the chronos #4, see Figure 12, one can notice the emergence of two peaks at $St = 0.21$ and $0.24$ and a broad region $0.20 - 0.23$ which should be attributed to the alternate vortex shedding. Another peak at the low frequency $St = 0.09$ seems to be also present, well immerged in the low frequency noise, which was previously shown in Zan's results (2008). On the whole, the signal remains quite noisy, especially at low frequencies: however, the emergence of the alternate vortex shedding is clearly detected. Due to the uncertainty in the wind velocity truly seen by the chimney, at least 5 % (or 10 % for the mean pressure coefficient), the Strouhal number must be taken with the same level of uncertainty.

The term #6 is an unsteady drag force due to the symmetrical shape of the topos. However, the corresponding chronos is so noisy that no remarkable frequency can be detected. Terms of higher rank progressively become disorganized, so that no physical meaning can be attributed to them.

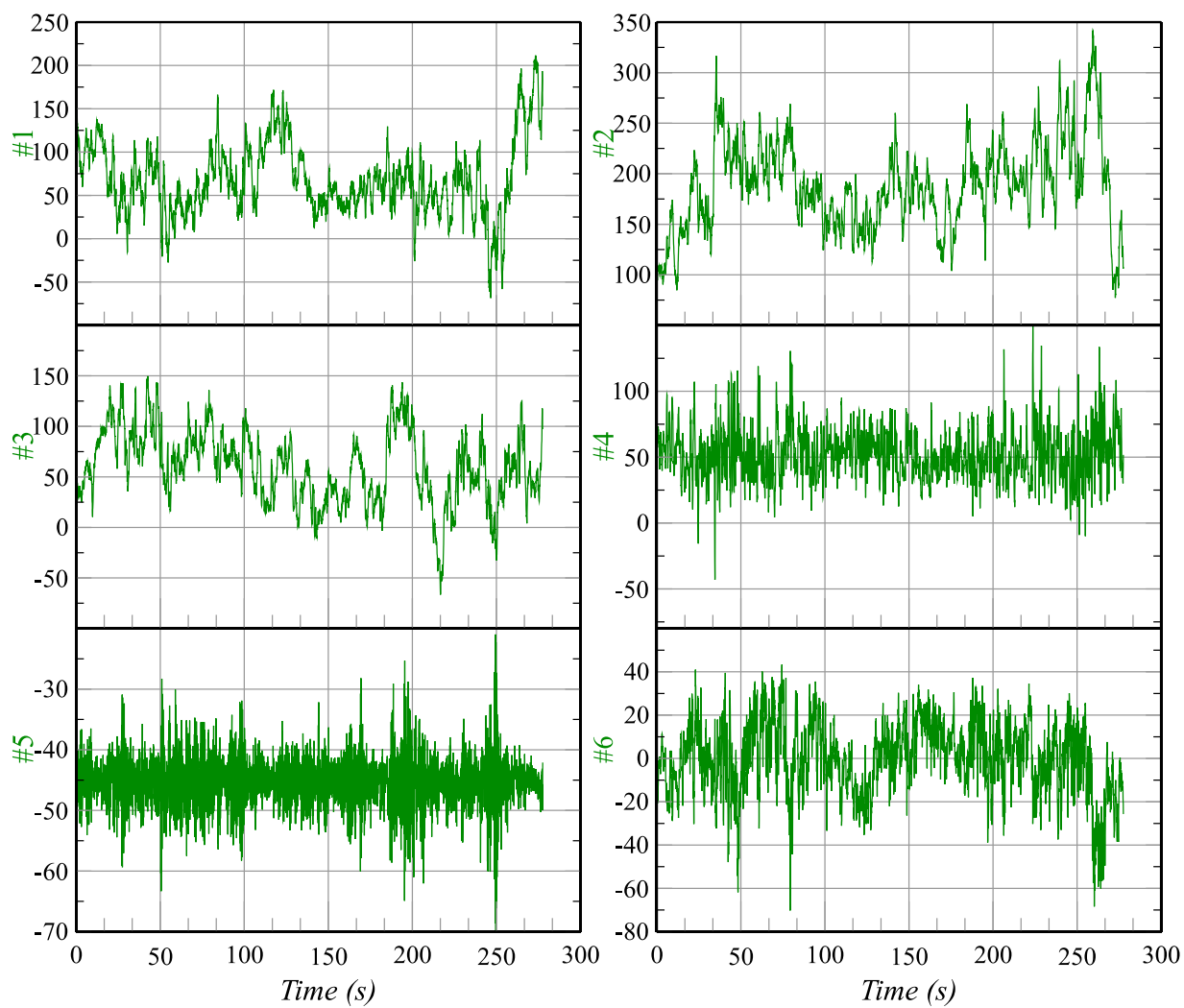

Figure 10. Six first chronos of the pressure distribution on the chimney.

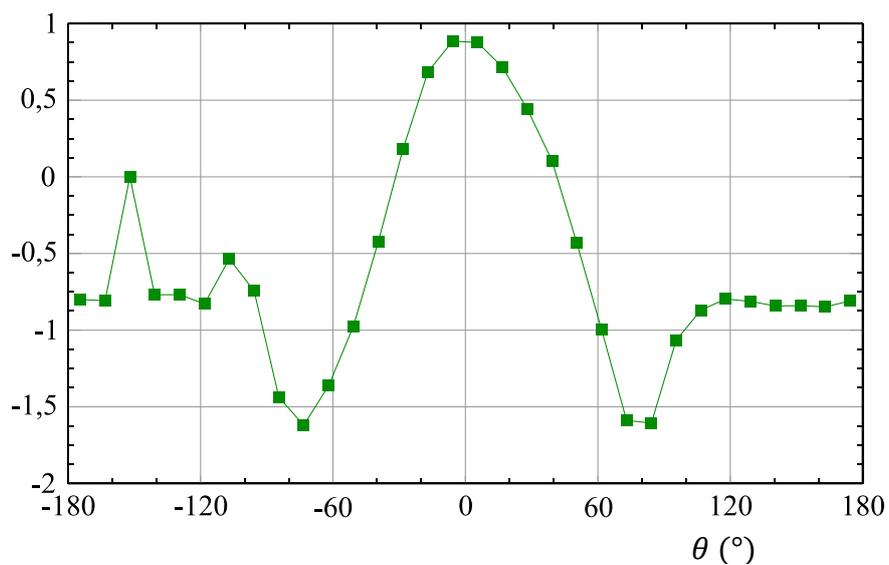

Figure 11. Mean pressure coefficient computed with the 3 first terms of the BOD following equation (11).

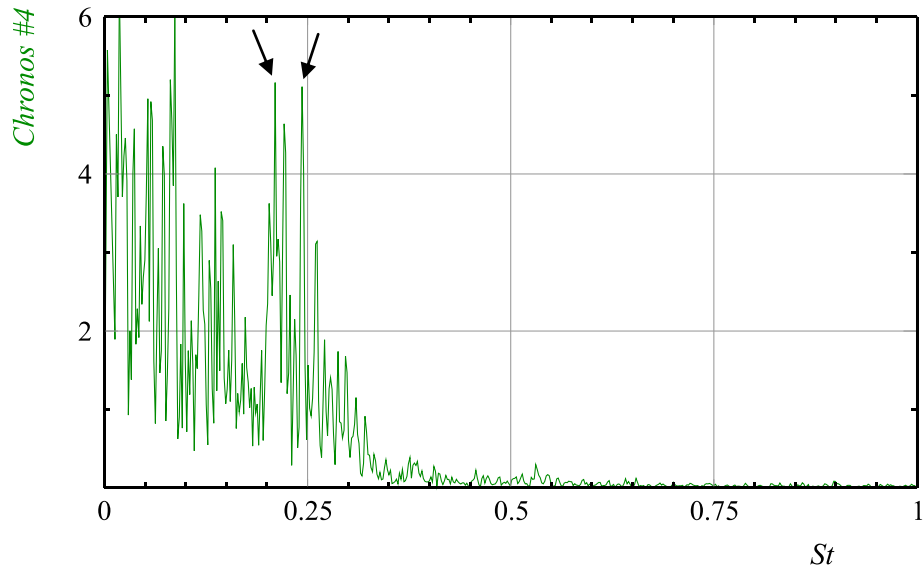

Figure 12. Spectral analysis of the chronos #4 showing the peaks at vortex shedding frequencies; quadratic average from 3 windows of 2048 samples, overlap 50%.

### 3.4. Comparison with the wind tunnel results

The comparison of the present measurements of the unsteady pressure distribution is made with two different wind tunnel experiments which were included in the project, both using 2D cylinder between walls: in the first, (Ellingsen et al. 2022a) the Reynolds number is fully maintained, but in a smooth flow, while the second (Ellingsen et al. 2022b) uses a turbulent flow and a small scale model. Therefore, the small Reynolds number is supposed to be compensated by added roughness at the surface of the cylinder. The Table 2 presents the main parameters of all these experiments.

For the small cylinder, the roughness is performed with ribs which produce a relative roughness $k/D = 0.0091$. The turbulent flow is generated by a grid upstream the test section. The resulting intensities are 6% in the direction of the flow and 3% for the other directions. The aspect ratio is large and provides good 2D conditions. However one should remember that the grid turbulence is very different from the atmospheric turbulence, obviously in terms of intensity level, but also in terms of length scales. In atmospheric flow the longitudinal length is much larger than the chimney diameter, while the length scale downstream a grid is of the order of the bar size that makes the grid, which in this case is close to the diameter of the cylinder. It known that both the turbulence intensity (Cheung & Melbourne 1983) and the length scale (Saathoff & Melbourne 1989) affect the unsteady loads and this will limit the scope of the current comparison.

Conversely, the smooth and large cylinder has an aspect ratio slightly lower than 10 between the walls of the wind tunnel, therefore it should be considered then as "almost 2D".

Measurements included synchronized wall pressure measurements of the same kind as for the chimney, which recording parameters were obviously adapted to the measurements conditions. All these details can be found in the corresponding previous references (Ellignsen et al. 2022a, 2022b).

Table 2. Main characteristics of the three experiments

| Object | Diameter $D$ (m) | Aspect ratio $L/D$ | $Re$ max | Inflow | Place |
|---|---|---|---|---|---|
| 2D rough cylinder | 0.055 | 36 | 66 000 | turbulent (grid) | Wind tunnel |
| 2D smooth cylinder | 0.5 | 9.5 | 2 200 000 | smooth | Wind tunnel |
| Chimney | 2 | 10 | 1 130 000 | atmospheric | Field |

Note that in wind tunnel, it was not possible to respect simultaneously the constraints of similarity of Reynolds number and the turbulent flow. Only one over two of these criteria is respected, and partially for the turbulent flow, while the chimney experiences both.

The comparison of the pressure coefficient, time-averaged and RMS is shown in Figure 13. Surprisingly the chimney, while at a supercritical Reynolds number, seems closer to the small rough cylinder in turbulent flow: especially the base coefficient is -0.8 in both cases. Moreover the minimum coefficient is -1.6 on the chimney against -1.2 on the small rough cylinder and -2.7 for the large smooth cylinder. This last data on the chimey agrees well with the one of (Kuniawati et al. 2022), despite the higher Reynolds number of the latter. Hence, the time averaged distribution of $\overline{Cp}$ of the chimney does not match nor the small scale wind tunnel 2D model, nor the 2D experiment at similar Reynolds number in smooth flow.

Concerning the RMS distribution, the turbulence of the flow acts as the most important mechanism. The levels measured on the chimney are much higher indeed, which should be linked to the upcoming wind. However, there are two remarkable peaks which emerge clearly at $\theta \approx \pm 100°$ and look well those observed on the large cylinder at $\theta \approx \pm 110°$. These peaks were identified as generated by the alternate vortex shedding, at the Strouhal number $St = 0.19$ for the cylinder (Ellingsen et al. 2022a). But for the chimney, the Strouhal number is more spread in the range $St = 0.20 - 0.24$, as previously found.

The comparison of the topos which generate the main unsteady lift due to the alternate vortex shedding shows, as for time-averaged pressure, that the chimney is closer to the small rough cylinder in turbulent flow, as it can be seen in Figure 14. This trend implies that the upstream turbulent flow has in fact a strong influence on the nature of the vortex shedding, greater than the Reynolds number similarity. In smooth flow at supercritical Reynolds number, the action of vortex shedding is concentrated on a small region of the circumference, with a high amplitude, while in turbulent flows, *ie* in wind tunnel or on site, the pressure fluctuations due to vortex shedding are spread along a larger surface along the circumference, on the two sides and the rear part.

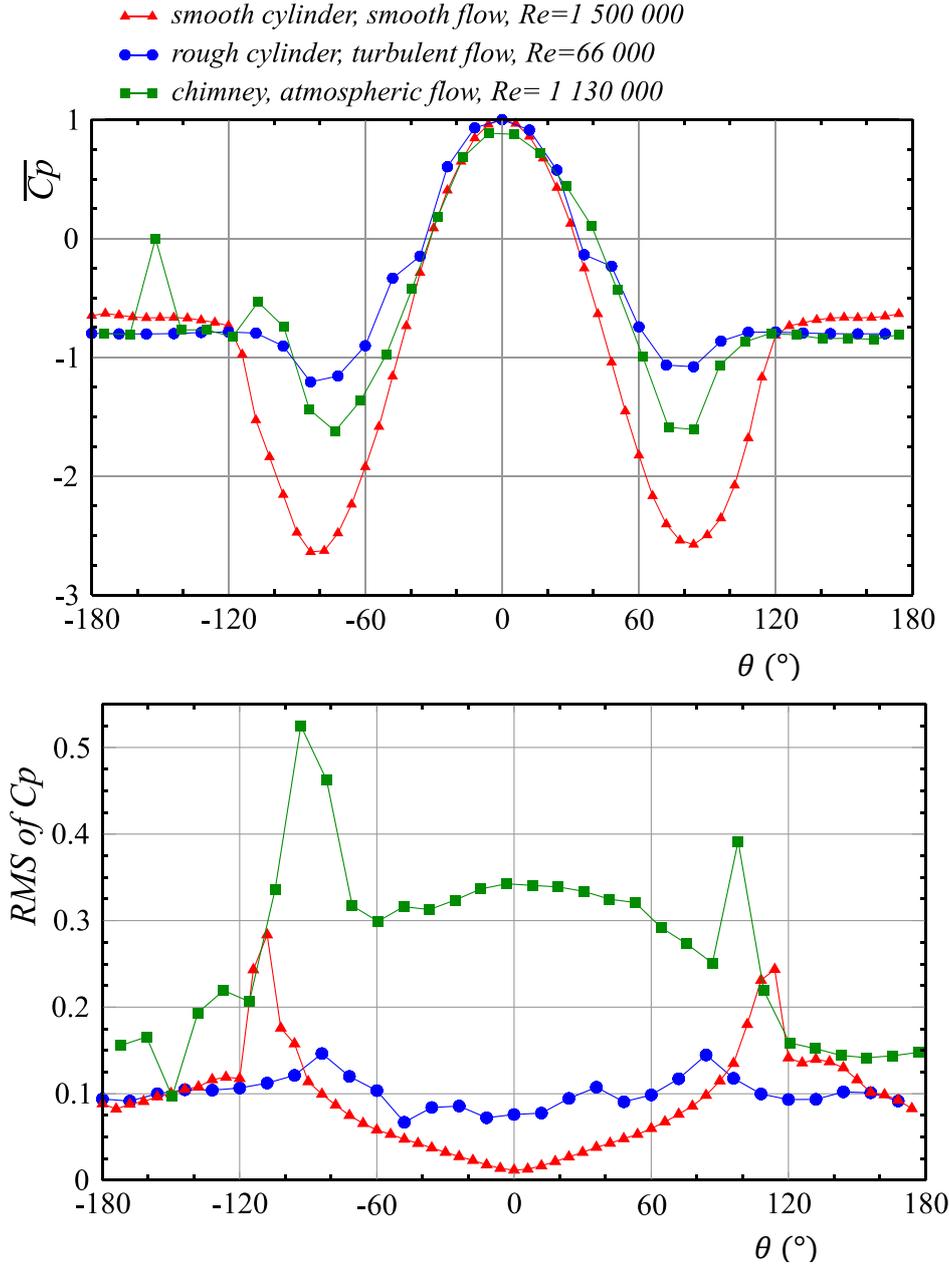

Figure 13. Comparison of the time-averaged (top) and RMS (bottom) pressure coefficients at the circumference of the cylinders

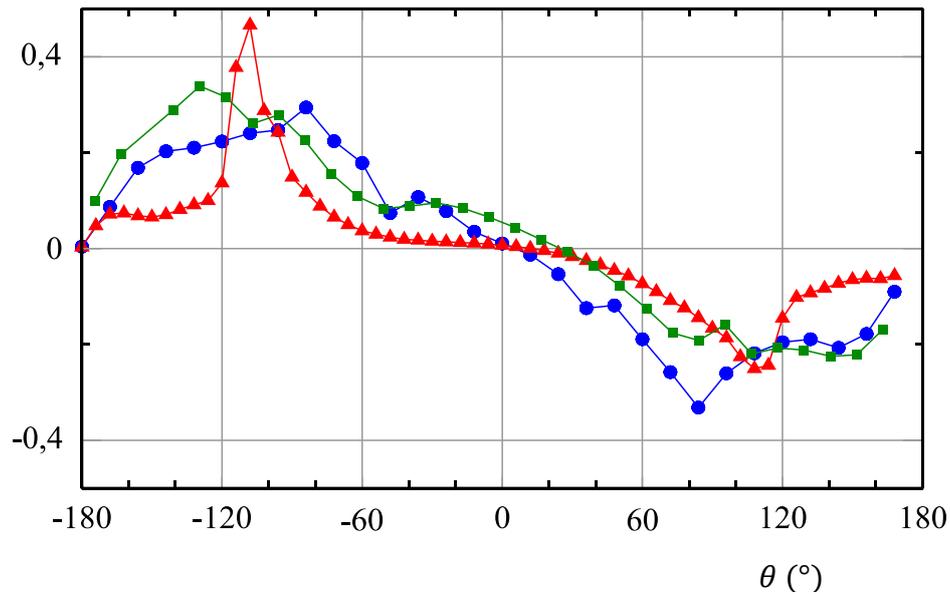

Figure 14. Comparison of the topos generating the lift force by vortex shedding. Same caption as in Figure 13.

## Summary and conclusion

We have presented the unsteady wall pressure measurements on a full scale 35 m circular chimney submitted to natural wind. The atmospheric turbulent boundary layer generates a turbulence intensity of 15 % at the altitude of the measurements where the corresponding Reynolds number is $1.13 \cdot 10^6$. The experiment is then in the so-called supercritical regime.

For the first time, the wall pressure distributions, steady and unsteady around the circumference, are presented despite large difficulties in the measuring process. The analysis of the pressure signals using the bi-orthogonal decomposition was really a key factor in the conduct of processing.

By comparison with wind tunnel similar measurements, the time-averaged pressure distribution is found between the one of a subcritical regime and a supercritical regime, slightly closer to the subcritical regime notably because of the base pressure coefficients that are similar.

The turbulence of the upstream flow is an essential ingredient for the unsteady pressure, which generates high levels of RMS pressure, much higher than all those measured in wind tunnel, even under turbulent flow.

The analysis by BOD allowed the extraction of a term which is generated by the alternate vortex shedding, which, while very noisy, the Strouhal number is found to be in the range 0.20-0.24. The corresponding shape of the pressure distribution (the topos) is compared with those found in wind tunnel: it is shown that the full scale field measurements are close to the small scale measurements in wind tunnel on a rough cylinder and under grid generated turbulent flow.

That said, a number of questions remain, for instance about the 3D character of the unsteady pressure along the height, which is essential in the chimney response, or the turbulent scales effects, that could not be studied here and require more investigations.

## Acknowledgements

This work results from the experiments included in a partnership co-funded by Beirens (Poujoulat Group), Centre Scientifique et Technique du Bâtiment (CSTB), Centre National d'Etudes Spatiales (CNES) and LadHyX, CNRS-Ecole polytechnique. Special acknowledgement is extended to Olivier Flamand from CSTB for the measuring systems and to Aurélien Jeanneton and François Coiffet from Beirens for the chimney design and erection. Photos of the chimney and drone operation were performed by Lilian Vezin.